\pdfoutput=1

\documentclass[preprint2]{aastex6}
\usepackage{apjfonts}

\usepackage{natbib}
\usepackage{amsmath}
\usepackage{amssymb}
\usepackage{subfigure}
\usepackage{url}
\usepackage{CJK}

\renewcommand{\vec}[1]{ {\mathbf #1} }

\newcommand{\Fig}{{Figure}}

\newcommand{\SDO}{{\it SDO}}
\newcommand{\Twm}{{$|T_w|_{\rm max}$~}}
\providecommand{\dodoi}[1]{doi:~\href{http://doi.org/#1}{\nolinkurl{#1}}}
\providecommand{\url}[1]{\href{#1}{#1}}
\providecommand{\doeprint}[1]{\href{http://ascl.net/#1}{\nolinkurl{http://ascl.net/#1}}}
\providecommand{\doarXiv}[1]{\href{https://arxiv.org/abs/#1}{\nolinkurl{https://arxiv.org/abs/#1}}}

\shorttitle{Study of Solar Magnetic Flux Ropes}
\shortauthors{Duan et al.}

\begin{document}
\begin{CJK*}{UTF8}{gbsn}

\title{A Study of Pre-flare Solar Coronal Magnetic Fields: Magnetic Flux Ropes}

\author{Aiying Duan \altaffilmark{1}$^*$,
  Chaowei Jiang \altaffilmark{2}$^*$,
  Wen He \altaffilmark{2},
  Xueshang Feng \altaffilmark{3},
  Peng Zou \altaffilmark{2},
  Jun Cui \altaffilmark{1,4}}

\altaffiltext{1}{School of Atmospheric Sciences, Sun Yat-sen University, Zhuhai 519000, China}


\altaffiltext{2}{Institute of Space Science and Applied Technology, Harbin Institute of Technology, Shenzhen 518055, China}

\altaffiltext{3}{SIGMA Weather Group, State Key Laboratory for Space Weather, National Space Science Center, Chinese
  Academy of Sciences, Beijing 100190, China}

\altaffiltext{4}{CAS Center for Excellence in Comparative Planetology, China}

\email{*Corresponding authors:}
\email{duanaiy@mail.sysu.edu.cn; chaowei@hit.edu.cn}

\begin{abstract}
  Magnetic flux ropes (MFRs) are thought to be the central structure
  of solar eruptions, and their ideal MHD instabilities can trigger
  the eruption. Here we performed a study of all the MFR
  configurations that lead to major solar flares, either eruptive or
  confined, from 2011 to 2017 near the solar disk center. The coronal
  magnetic field is reconstructed from observed magnetograms, and
  based on magnetic twist distribution, we identified the MFR, which
  is defined as a coherent group of magnetic field lines winding an
  axis with more than one turn. It is found that 90\% of the events
  possess pre-flare MFRs, and their three-dimensional structures are
  much more complex in details than theoretical MFR models. We further
  constructed a diagram based on two parameters, the magnetic twist
  number which controls the kink instability (KI), and the decay index
  which controls the torus instability (TI). It clearly shows lower
  limits for TI and KI thresholds, which are $n_{\rm crit} = 1.3$ and
  $|T_w|_{\rm crit} = 2$, respectively, as all the events above
  $n_{\rm crit}$ and nearly 90\% of the events above
  $|T_w|_{\rm crit}$ erupted. Furthermore, by such criterion, over
  70\% of the events can be discriminated between eruptive and
  confined flares, and KI seems to play a nearly equally important
  role as TI in discriminating between the two types of flare. There
  are more than half of events with both parameters below the lower
  limits, and 29\% are eruptive. These events might be triggered by
  magnetic reconnection rather than MHD instabilities.
\end{abstract}

\keywords{Magnetic fields;
          Magnetohydrodynamics (MHD);
          Methods: numerical;
          Sun: corona;
          Sun: flares}

\section{Introduction}
\label{sec:intro}

As a leading cause of space weather, solar eruptions, including flares
and coronal mass ejections (CMEs) are still difficult to predict. Now
it is commonly believed that solar eruptions have their root in the
evolution of magnetic field in the solar atmosphere. In particular,
the magnetic field dominates the dynamics in the solar corona, which
is a highly electric-conducting plasma environment. However, two key
questions arise in understanding the physics of solar eruptions: what
is magnetic structure of the corona before eruption and what is the
triggering mechanism of the eruption?  Observations show that major
flares are often associated with CMEs, but many flares do not, which
are named as confined flares. Thus another important question is what
is the factor that determines such difference?

Through several decades of studies, a variety of models have been
proposed to answer these questions~\citep[e.g., see review papers of
][]{Forbes2006, Shibata2011, Aulanier2014, Schmieder2012, Schmieder2013,
  Janvier2015}. In a rough classification, these models fall into two
categories, one is based on magnetic reconnection~\citep{Mikic1994,
  Antiochos1999, Moore2001} and the other is ideal MHD
instabilities~\citep{Bateman1978, Hood1981, Torok2004, Torok2005, Kliem2006,
  Fan2007, Aulanier2010}. In the former group, there are two models
most frequently invoked, namely the run-away tether-cutting
reconnection~\citep{Moore2001} and the breakout
reconnection~\citep{Antiochos1999}. Both these models assume the
pre-flare magnetic field as a strongly sheared configuration with a
topology prone to reconnection, and an eruption will occur if a
positive feedback between the reconnection and the outward expansion
of the sheared magnetic flux can be established. In the tether-cutting
model, internal reconnection between the sheared magnetic arcades
triggers the eruption, while in the breakout model, reconnection takes
place externally, above or aside of the sheared arcades in a magnetic
null point topology. However, in what conditions the feedback can be
triggered is still elusive.

In the ideal MHD models, two kinds of ideal instabilities are most
extensively investigated in the context of solar eruptions, which are
the helical kink instability~\citep[KI, ][]{Hood1981, Torok2004,
  Torok2005, Torok2010} and the torus instability~\citep[TI,
][]{Kliem2006}. Both of the two instabilities are developed based on a
fundamental magnetic configuration in the plasma known as magnetic
flux rope (MFR)~\citep{Kuperus1974, Chen1989, Titov1999,
  Amari2014nat}, which is a coherent group of twisted magnetic flux
that winding around a common axis. Naturally in models of the ideal
MHD instabilities, MFR must exist in the corona prior to an eruption,
and evolution in the photosphere can then, often slowly, build up the
MFR to an unstable regime and produce the eruption. From a theoretical
point of view, the existence of pre-eruptive MFRs should be common in
the corona. This is because the coronal magentic field is
approximately force-free such that the electric currents direct
dominantly along magnetic field lines, thus such field-aligned
currents introduce poloidal magnetic flux around the currents, which
has a potential to make the field lines twist and form MFRs. Indeed,
the relevance of MFRs with solar flares and eruptions has also been
extensively evidenced from both observations and coronal magnetic
field reconstructions. For instance, X-ray and EUV sigmoid, filament,
EUV hot channel, and coronal cavity are invoked as indirect
observations of coronal MFRs~\citep[e.g., see a recent review paper by
][]{ChengX2017}. Using nonlinear force-free field (NLFFF)
extrapolations from vector magnetograms, which is a basic tool for
unraveling the 3D information of solar coronal magnetic field, MFRs
were identified frequently~\citep[e.g., see another recent review
by][]{GuoY2017}.

The trigger of MFR eruption can be through either KI or TI, basing
mainly on two critical parameters. KI is controlled by the twist
degree of MFR, occurring if the twist of the MFR exceeds a critical
value. Through a eruptive expansion, KI will transform the excessive
magnetic twist to a writhe through a helical deformation of the MFR
axis. Theoretical and numerical investigations have shown that the KI
threshold, as measured by the winding number of magnetic field
  lines around the MFR's axis, seems to have a wide range from
$\sim 1.25$ turns to $\sim 2.5$ turns~\citep{Baty2001, Fan2003,
  Torok2003, Torok2004, Torok2005}, which depends on the details of
the MFR, such as the geometry of the axis, the aspect ratio of the MFR
(i.e., ratio of length of the rope to the size of its cross section),
and the line-tying effects by the photosphere. On the other hand, TI
is controlled by the decay index~\citep{Torok2005, Torok2007,
  Kliem2006}, which is the spatial deceasing speed of the MFR's
overlying magnetic field that strapping the MFR. Assuming a outward
quasi-static expansion of the MFR due to its hoop force, which is
resulted by the self-inductive effect of the current in the MFR, both
the hoop force and the strapping force will decrease with the
expansion. If the strapping force decreases faster than the hoop
force, the system will be unstable because the net force points to the
direction of expansion. The TI threshold of decay index is found to
have typical values in the domain of $1.1 \sim 1.7$, again derived
from a series of theoretical and numerical
investigations~\citep{Kliem2006, Torok2007, Fan2007, Aulanier2010,
  Demoulin2010, Fan2010, Olmedo2010, Zuccarello2015}. An attractive
advantage of these ideal MHD models is that the controlling
parameters, e.g., the twist degree of the MFR and the decay index of
the strapping field, can be used potentially in forecasting the
eruptiveness of flares.

The theory of KI and TI has been recently, and is becoming even more
widely, applied to study real solar eruptions. Case studies of solar
eruptions using NLFFF reconstructions often shows MFRs exist prior to
flare and its ensuing eruption is very likely due to
KI~\citep[e.g.,][]{LiuR2016} or TI~\citep[e.g.,][]{ChengX2013,
  JiangC2013MHD, JiangC2018}, as the reconstructed MFRs appears to be
close to the thresholds. However, due to the intrinsic complexity of
the magnetic field in the solar corona, the configuration of MFRs can
be very different from case to case, and cannot be fully characterized
by the KI and TI theory, which are both based on relatively simplified
or idealized configuration. Furthermore, a recent laboratory
experiment of MFRs emulating the dynamic behaviors of solar line-tied
MFRs suggests that the theory might miss including the magnetic
tension force caused by the toroidal magnetic flux of the
rope~\citep{Myers2015}. If such magnetic tension is strong enough,
i.e., the toroidal flux is large enough (and thus corresponding to a
sufficiently small magnetic twist), it can restrict the flux rope from
eruption even it fulfills the TI condition, for which the authors call
it as a `failed torus' event.  Thus the application of the theoretical
parameters of KI and TI is still not straightforward.

A very recent statistic study of the controlling parameters of KI and
TI for solar flares was performed by~\citet{JingJ2018}. They surveyed
38 major flares, including 26 ejective and 12 confined ones, by NLFFF
reconstructions of the pre-flare coronal magnetic field using a code
developed by~\citet{Wiegelmann2004}. Then for each events, the
reconstructed 3D magnetic field is analyzed by computing the magnetic
twist and decay index. It was found that the KI parameter, i.e., the
twist number appears to play no role in discriminating between the
confined and eruptive events. And for the TI parameter, the threshold
of decay index is found to be $\sim 0.75$, which is much lower than
the typical values that are derived in theoretical and numerical
studies. However, as also pointed out by the authors, such results
might strongly depends on the quality or reliability of the coronal
magnetic field reconstructions. Currently there are many methods
available for NLFFF extrapolations from the vector magnetograms, but
different methods seem to produce rather inconsistence results between
each other~\citep[e.g.,][]{DeRosa2009, Regnier2013, Aschwanden2014,
  DuanA2017, Wiegelmann2017}. Thus any results based on any single
NLFFF code must be taken with cautions, and independent studies with
different codes are required for a better inspection. Thus, one of the
purposes of this paper is to see how the results behave if using an
independent NLFFF code to perform a similar statistical
investigation. The other purpose is, for the first time, to
statistically investigate the complexity of the pre-flare coronal
MFRs.

In this paper, we employed the coronal magnetic field reconstruction
method developed by \citet{JiangC2013NLFFF}, named the CESE--MHD--NLFFF
code, to study a slightly larger sample of 45 major flares with 29
eruptive and 16 confined. We attempt to reveal the complexity of MFRs
in solar corona by showing the magnetic configuration of each
MFR. With a much stricter definition of MFR and a more relevant way of
calculating the decay index, our study shows that the KI and TI parameters
play an equal important role in discriminating between the eruptive
and confined events, and the TI thresholds for the eruptive events
is much closer to the theory. The rest of the paper is organized as
follows: Data and method are presented in Section~\ref{sec:method},
then results are given in Section~\ref{sec:res}, and finally
discussions and conclusions are made in Section~\ref{sec:concl}.

\begin{table*}[htbp]
\footnotesize
  \centering
  \caption{List of events and properties of their MFRs.}
  \begin{tabular}{cccccccc}
    \hline
    \hline
    No. & Flare peak time & Flare class  & NOAA AR & Position & E/C$^{a}$ & $T_w$ & $n$  \\
    \hline
    1 & SOL2011-02-13T17:38 & M6.6 & 11158 & S20E04 & E &  0.76 & 0.99 \\
    2 & SOL2011-02-15T01:56 & X2.2 & 11158 & S20W10 & E &  1.52 & 0.98 \\
    3 & SOL2011-03-09T23:23 & X1.5 & 11166 & N08W09 & C & -1.75 & 0.50 \\
    4 & SOL2011-07-30T02:09 & M9.3 & 11261 & S20W10 & C & -0.88 & 0.51 \\
    5 & SOL2011-08-03T13:48 & M6.0 & 11261 & N16W30 & E &  2.45 & 1.40 \\
    6 & SOL2011-09-06T01:50 & M5.3 & 11283 & N14W07 & E &  0.92 & 0.52 \\
    7 & SOL2011-09-06T22:20 & X2.1 & 11283 & N14W18 & E &  1.02 & 1.65 \\
    8 & SOL2011-10-02T00:50 & M3.9 & 11305 & N12W26 & C & -0.92 & 0.42 \\
    9 & SOL2012-01-23T03:59 & M8.7 & 11402 & N28W21 & E & -1.63 & 0.73 \\
    10 & SOL2012-03-07T00:24 & X5.4 & 11429 & N17E31 & E & -2.11 & 0.71 \\
    11 & SOL2012-03-09T03:53 & M6.3 & 11429 & N15W03 & E & -1.17 & 0.73 \\
    12 & SOL2012-05-10T04:18 & M5.7 & 11476 & N12E22 & C & -1.11 & 1.21 \\
    13 & SOL2012-07-02T10:52 & M5.6 & 11515 & S17E08 & E & -1.56 & 0.35 \\
    14 & SOL2012-07-05T11:44 & M6.1 & 11515 & S18W32 & C &  1.14 & -0.41 \\
    15 & SOL2012-07-12T16:49 & X1.4 & 11520 & S15W01 & E &  2.20 & 0.42 \\
    16 & SOL2013-04-11T07:16 & M6.5 & 11719 & N09E12 & E & -1.10 & 0.26 \\
    17 & SOL2013-10-24T00:30 & M9.3 & 11877 & S09E10 & E &  2.00 & 0.56 \\
    18 & SOL2013-11-01T19:53 & M6.3 & 11884 & S12E01 & C &  1.50 & 0.42 \\
    19 & SOL2013-11-03T05:22 & M4.9 & 11884 & S12W17 & C &  3.00 & 0.07 \\
    20 & SOL2013-11-05T22:12 & X3.3 & 11890 & S12E44 & E &  1.35 & 2.72 \\
    21 & SOL2013-11-08T04:26 & X1.1 & 11890 & S12E13 & E &  1.26 & 1.87 \\
    22 & SOL2013-12-31T21:58 & M6.4 & 11936 & S15W36 & E & -2.20 & 1.11 \\
    23 & SOL2014-01-07T10:13 & M7.2 & 11944 & S13E13 & C &  1.65 & 0.21 \\
    24$^{*}$ & SOL2014-01-07T18:32 & X1.2 & 11944 & S15W11 & E &  6.50 & 0.20 \\
    25 & SOL2014-02-02T09:31 & M4.4 & 11967 & S10E13 & C & -1.73 & -0.12 \\
    26 & SOL2014-02-04T04:00 & M5.2 & 11967 & S14W06 & C & -1.90 & 1.03 \\
    27 & SOL2014-03-29T17:48 & X1.1 & 12017 & N10W32 & E &  1.53 & 1.72 \\
    28 & SOL2014-04-18T13:03 & M7.3 & 12036 & S20W34 & E &  2.30 & 1.82 \\
    29 & SOL2014-09-10T17:45 & X1.6 & 12158 & N11E05 & E & -0.85 & 0.17 \\
    30$^{*}$ & SOL2014-09-28T02:58 & M5.1 & 12173 & S13W23 & E & -2.76 & 1.96 \\
    31 & SOL2014-10-22T14:28 & X1.6 & 12192 & S14E13 & C & -1.10 & 0.94 \\
    32 & SOL2014-10-24T21:41 & X3.1 & 12192 & S22W21 & C & -1.79 & 0.64 \\
    33 & SOL2014-11-07T17:26 & X1.6 & 12205 & N17E40 & E &  3.55 & 1.21 \\
    34 & SOL2014-12-04T18:25 & M6.1 & 12222 & S20W31 & C &  2.60 & 0.60 \\
    35 & SOL2014-12-17T04:51 & M8.7 & 12242 & S18E08 & E &  0.70 & 0.66 \\
    36 & SOL2014-12-18T21:58 & M6.9 & 12241 & S11E15 & E &  1.09 & 1.49 \\
    37 & SOL2014-12-20T00:28 & X1.8 & 12242 & S19W29 & E &  1.32 & 0.56 \\
    38 & SOL2015-03-11T16:21 & X2.1 & 12297 & S17E22 & E &  2.04 & 1.80 \\
    39 & SOL2015-03-12T14:08 & M4.2 & 12297 & S15E06 & C &  1.10 & 0.72 \\
    40 & SOL2015-06-22T18:23 & M6.5 & 12371 & N13W06 & E & -1.24 & 1.51 \\
    41 & SOL2015-06-25T08:16 & M7.9 & 12371 & N12W40 & E & -2.90 & 0.49 \\
    42 & SOL2015-08-24T07:33 & M5.6 & 12403 & S14E00 & C &  1.04 & 0.33 \\
    43 & SOL2015-09-28T14:58 & M7.6 & 12422 & S20W28 & C & -1.25 & 1.20 \\
    44 & SOL2017-09-04T20:33 & M5.5 & 12673 & S10W11 & E & -1.43 & 1.09 \\
    45 & SOL2017-09-06T12:02 & X9.3 & 12673 & S09W34 & E & -1.80 & 1.72 \\
    \hline
  \end{tabular}
  \tablenotetext{a}{E--eruptive, C--confined.}
  \tablenotetext{*}{Event 24 occurred between NOAA ARs 11944 and
    11943, and event 30 occurred between ARs 12173 and 12172.}
  \label{tab:event_list}
\end{table*}

\section{Method}
\label{sec:method}

\subsection{Event Selection}
Since we are interested in the major flares for which coronal magnetic
field extrapolation can be perform with reliable observed
magnetograms, we use the similar criterion for selecting events
samples as employed in~\citet{Toriumi2017} and~\citet{JingJ2018}. That
is all the flares above GOES-class M5 (in general) that occurred
within 45 degree of the solar disk center from 2011 January to 2017
December, and most of them occurred in active regions
  (ARs). For the confined flares, the flare class criterion is
relaxed to include also M3.9. Furthermore, since some ARs (e.g.,
AR12673) produced several (more than three) flares fulfilling the
above criterion, we select only two flares to avoid the
over-representation for a certain AR: the first one is the largest
flare, and the second one is the flare occurring nearest to the disk
center. But if these two flares are both eruptive and meanwhile the AR
also produced one or more confined flares, we replace the second one
with the confined flare (the largest one if there are more than one
confined flares). There are 45 events in total, including 29 eruptive
flares and 16 confined flares from 30 different ARs, as listed in
Table~\ref{tab:event_list}. Note that two events (number 24 and 30)
are inter-AR flares.

\subsection{Coronal Magnetic Field Reconstructions}
For each event, we carried out 3D magnetic field reconstruction for
the pre-flare corona from the {\SDO}/HMI vector magnetograms using the
CESE--MHD--NLFFF code~\citep{JiangC2013NLFFF}. The last available
magnetogram for at least 10 minutes before the flare GOES start time
is used to avoid the possible artifacts introduced by the strong flare
emission. In particular, we used the data product of the Space-weather
HMI Active Region Patch \citep[SHARP,][]{Bobra2014}, in which the
180$^{\circ}$ ambiguity has been resolved by using the minimum energy
method, the coordinate system has been modified via the Lambert
method, and the projection effect has been corrected. The
CESE--MHD--NLFFF model is based on an MHD-relaxation method which
seeks approximately force-free equilibrium. It solves a set of
modified zero-$\beta$ MHD equations with a friction force using an
advanced conservation-element/solution-element (CESE) space-time
scheme on a non-uniform grid with parallel computing
\citep{JiangC2010}. The code also utilizes adaptive mesh refinement
and a multi-grid algorithm to optimize the relaxation process. This
model has been tested by different benchmarks including a series of
analytic force-free solutions \citep{Low1990} and numerical MFR models
\citep{Titov1999}. The results of extrapolation reproduced from
\textit{SDO}/HMI are in good agreement with corresponding observable
features like filaments, coronal loops, and sigmoids
\citep{JiangC2013NLFFF, JiangC2014NLFFF}.

\subsection{Magnetic Twist Number and Identification of MFRs}
It is nontrivial to identify MFRs in a reconstructed coronal
  magnetic field because their configuration are generally complex
  compared with theoretical models. Here the search of MFR is based on
  the distribution of a parameter called magnetic twist
  number~\citep{Berger2006}, which can be conveniently computed
  without resorting to the geometry of an MFR~\citep{LiuR2016}. The
magnetic twist number $T_w$ for a given (closed) field line is defined
as
\begin{equation}\label{Tw}
  T_w=\int_L \frac{(\nabla \times \vec B)\cdot \vec B}{4\pi B^2} dl
\end{equation}
where the integral is taken along the length $L$ of the magnetic field
line from one footpoint on the photosphere to the other. As
  shown by \citet{LiuR2016}, $T_w$ provides an good approximation of
  the number of turns that two infinitesimally close field lines wind
  about each other. Thus $T_w$ is not identical to the classic winding
  number of field lines about a common axis, the parameter often
  used in the analysis of the helical KI. Nevertheless,
according to \citet{LiuR2016}'s analysis, the magnetic field line that
possesses the extremum value (maximum or minimum) of $|T_w|$
in an MFR can be reliably regarded as the rope axis, and
  $T_{w}$ computed in the vicinity of the axis approaches the winding
  number. For each 3D reconstructed magnetic field data, we compute
the twist number on grid points with a resolution 4 times of the
original data, from which a 3D smooth distribution of $T_w$ is
obtained. Basing on this distribution of $T_w$, the MFR in the field
can be precisely identified.

\begin{figure*}
  \centering
  \includegraphics[width=0.9\textwidth]{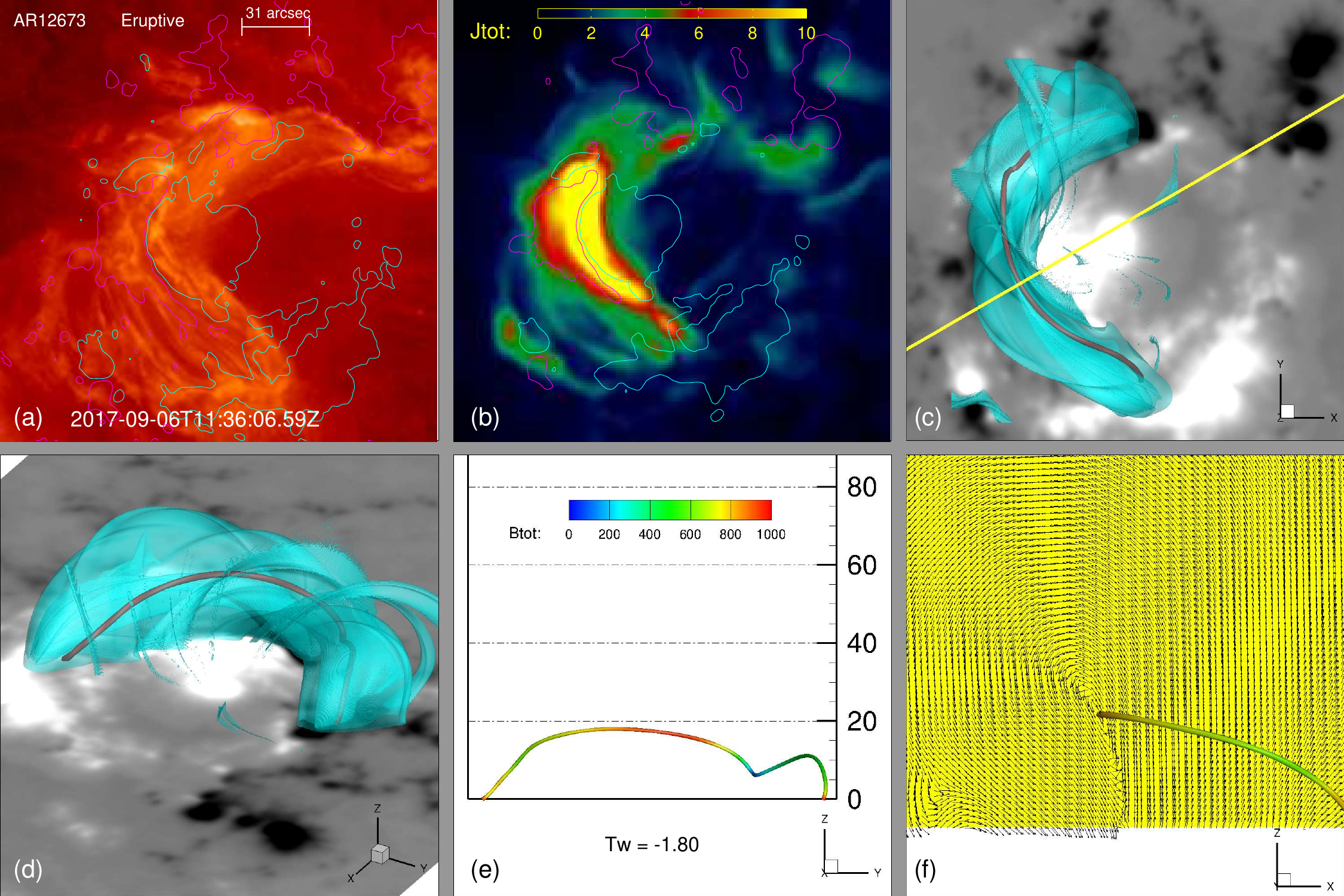}
  \caption{An example showing the identification of MFR. The figure
    shows the 45th event, i.e., the pre-flare magnetic field of the
    X9.3 flare in AR~12673. (a) {\SDO}/AIA~304~{\AA} image of the
    pre-flare corona. (b) A vertical integration of the current
    density derived from the reconstructed magnetic field, showing the
    strong-current region. (c) The cyan, transparent object is the
    iso-surface of the $|T_w|=1$ and the red, thick line represent the
    magnetic field line possessing the maximum value of $|T_w|$, which
    is regarded as the axis of the MFR. (d) The same structure of (c)
    but in a different angle of view in 3D. (e) Side view of the rope
    axis with colors denote the magnetic field strength on the
    line. The unit of the $z$ axis is 1~arcsec (or 720km).  (f) A
    central vertical cross section of the MFR whose location is dented
    by the yellow line in panel (c), with the transverse field on the
    slice shown by the arrows, which forms spirals centered at the
    axis of the rope denoted by the thick line.}
  \label{MFR_example}
\end{figure*}

There is no accurate definition of solar coronal MFR in the
literature. Generally, an MFR refers to a group of magnetic field
lines spiraling around the same axis with certain twist, but there
seems to be no consensus on what extent of the twist degree can be
regarded as a rope. Here we follow the definition of \citet{LiuR2016},
that is, the MFR is defined as a coherent group of magnetic field
lines with $|T_w| \ge 1$ (i.e., field lines spiral above a full turn)
and with the same sign of twist. The coherence means that the volume
of the magnetic flux with $|T_w| \ge 1$ forms a single tube without
segmentation. In some events, there are more than one MFR as the
magnetic flux with $|T_w| \ge 1$ forms multiple, separate tubes, and
even more complex, different MFRs can have inverse signs of twist from
each other (which will be shown in Section~\ref{sec:res}), indicating
the intrinsic complexity of the coronal magnetic field. There are
places where the $|T_w|$ is strong, but forms a sheet like structure
with width close to the grid size. These structures are actually
complex magnetic separatrix layers or quasi-separatrix layers and
cannot be defined as MFR, although their twist numbers appear
large. The reason which makes the $|T_w|$ high is the relatively high
value of $J/B$ in these regions. We exclude these regions in searching
flux rope. Furthermore, the search of MFR is aid by {\SDO}/AIA
observations such that the MFR is restricted within the flare site,
and especially the morphology of the MFR is compared with pre-eruptive
filaments if they are observable. As an example, in
\Fig~\ref{MFR_example}, we show the MFR in the 45th event, which is
the pre-flare magnetic field of the largest X-class flare in AR~12673,
also the largest one of solar cycle 24~\citep{SunX2017, Inoue2018ApJ,
  LiuL2018, Mitra2018, Seaton2018, WangH2018, Getling2019,
  Petrie2019}. Clearly the MFR is identified by a volumetric channel
with $|T_w| \ge 1$ and strong current density. The MFR body appears
roughly to be a C shape following the main PIL of the AR.

\subsection{KI Parameter}
By a comprehensive study of the magnetic twist distribution and
evolution in an AR that produced a series of flares, \citet{LiuR2016}
suggests that the magnetic field line with maximum twist number in
their studied MFR is a reliable proxy of the rope
axis. Further, they found that, comparing to other parameters like
magnetic energy and helicity, the maximum twist number changes most
prominently across the flares, as it increases systematically before
each flare and decrease stepwise after it.  This suggests that the
MFR's maximum twist number, \Twm, is very sensitive in association
with KI occurring in flares. We thus employed the \Twm in our analysis
as the KI controlling parameter.

After locating the MFRs, we can then locate their axis, which is
defined as a single field line that possesses the maximum value of the
twist number, \Twm, in the MFR. As shown in \Fig~\ref{MFR_example},
the axis of the rope is the field lines with largest twist of
$|T_w|=1.8$. It is fully wrapped by the iso-surface of $|T_w|=1$,
running horizontally in the central part and reaching a height of
roughly $20$~arcsec. The location of the axis is double checked by
using vertical slices cutting through the rope axis in a perpendicular
direction and to see if the poloidal flux of the rope forms rings
centered at the axis in such cross section, as shown in
\Fig~\ref{MFR_example}(f). This is fulfilled for most of the events,
suggesting that the field line with maximum $T_w$ is a reliable proxy
of the rope axis, which is consistent with the findings
of~\citet{LiuR2016}.\footnote{It should be noted that there
    are cases in which the MFR axis is located at a local minimum of
    $|T_{w}|$, especially in decayed active regions and the quiet
    Sun~\citep[e.g.,][]{Su2011}.}  For those events that possess
multiple MFRs, axis for each one can be identified
independently. There are complex cases in which the field line with
\Twm in the MFR can runs partly on the surface of the rope, because of
the unevenly distribution of the twist number.

\begin{figure*}[htbp]
  \centering
  \includegraphics[width=0.9\textwidth]{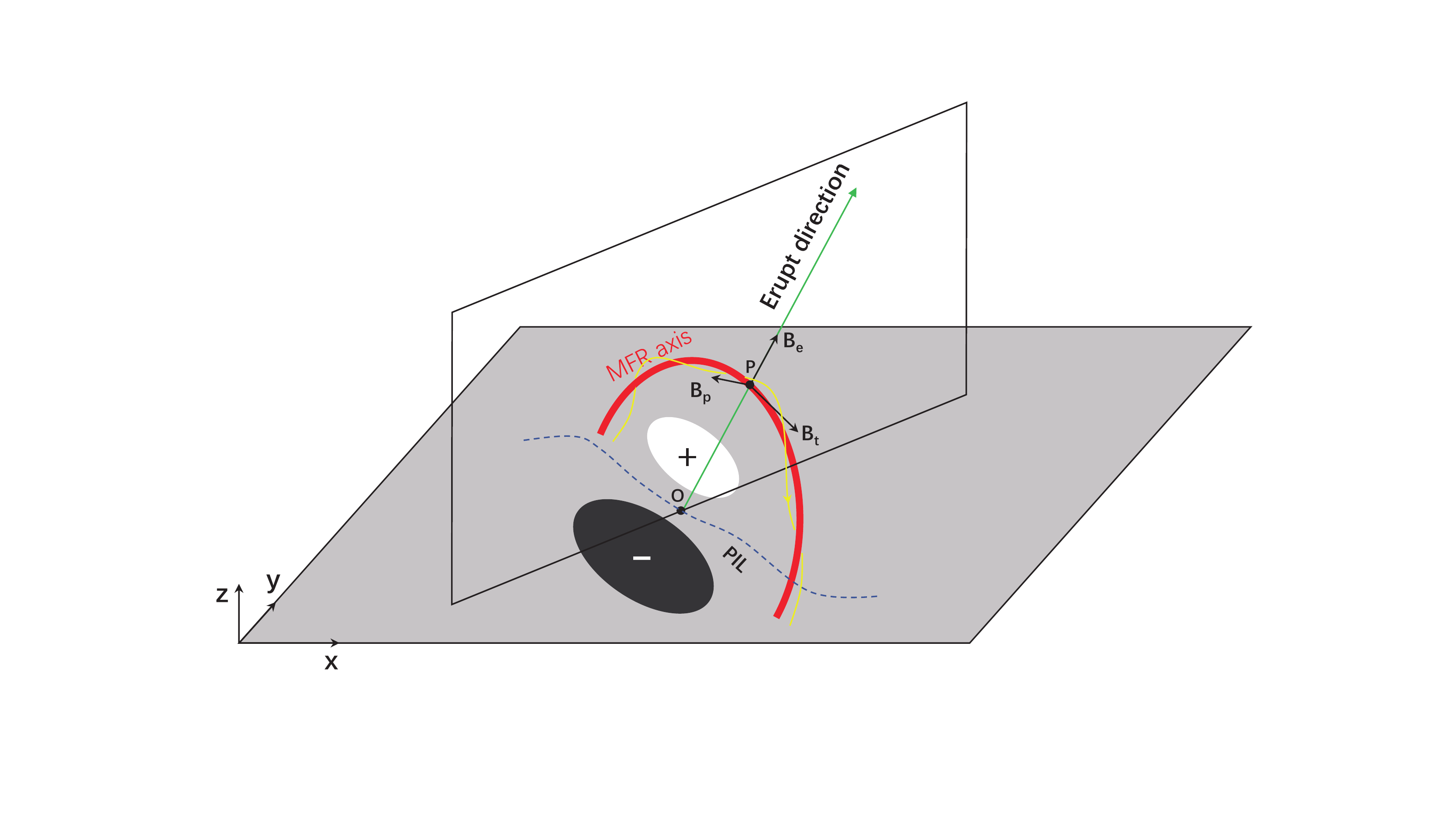}
  \caption{Illustration of calculating the decay index $n$ at the apex
    of the MFR axis. The thick red curve is the MFR axis, and P
    denotes the highest point along the axis. A vertical slice cutting
    perpendicularly through the axis at P. The intersection of the
    vertical slice with the main PIL is marked as O. The background
    magnetic field $\vec B$, which is computed based on the potential
    field model, are decomposed into three components as $B_e$, $B_p$,
    and $B_t$. }
  \label{oblique_decay_index}
\end{figure*}

\subsection{TI Parameter}
In many literatures, the decay index is simply defined as
\begin{equation}
n = -\frac{d \log(B)}{d \log(h)}
\end{equation}
where $B$ denotes the strapping field stabilizing the MFR and $h$ is
the vertical height locally or radial distance globally, assuming that
the MFR erupts vertically or radially and the strapping force point in
the opposite direction. However, since the triggering and initiation
of MFR eruption is strongly influenced by its complex magnetic
environment in the lower corona, which is often non-symmetric with
respect to the PIL such that the eruption direction is not along the
vertical (or radial) direction. Such non-radial eruptions are
frequently observed in filament eruptions~\citep{McCauley2015}. In
such case, computing the decay index along an oblique line matching
the eruption direction ought to be more accurate. As illustrated in
\Fig~\ref{oblique_decay_index}, the MFR is significantly inclined to the right side away from the
vertical direction, owing to the stronger magnetic flux distribution (and
thus the magnetic pressure) in the left side. To define the decay
index in such configuration, we first use a vertical slice cutting
through the middle of the rope, generally, at the apex of the axis
(i.e., highest point on the axis, marked as P in the \Fig), in
perpendicular direction to the rope axis. The intersection point of
the bottom PIL with the slice is marked as O as shown in the
\Fig. Then, we calculate decay index in the OP direction with O as the
starting point. As usual, the potential field model extrapolated from
the $B_z$ component of the photospheric magnetogram can be considered
as a good approximation of the external (strapping) magnetic field
with respect to the MFR. To be more relevant, we further decompose the
potential field into three orthogonal components $B_e$, $B_p$, and $B_t$, where
$B_e$ is along OP, $B_p$ is perpendicular to the OP on the slice,
and $B_t$ is perpendicular to the slice. In defining $n$ we
only use the poloidal flux $B_p$, that is
 \begin{equation}
n = -\frac{d \log(B_p)}{d \log(r)}
\end{equation}
(where $r$ is the distance pointing from O to P). This is because the cross
product of the current of the rope (which is along the axis) with only
the poloidal flux of the overlying field can produce the strapping
force (directing P to O) that stabilizing the MFR. The other
components, for example, $B_t$ is parallel to the current and has no
effect, and the cross product of current with $B_e$ produce a force
parallel to the $B_p$ which only controls the eruption direction and
furthermore $B_e$ is often small.

\begin{figure*}[htbp]
  \centering
  \includegraphics[width=0.8\textwidth]{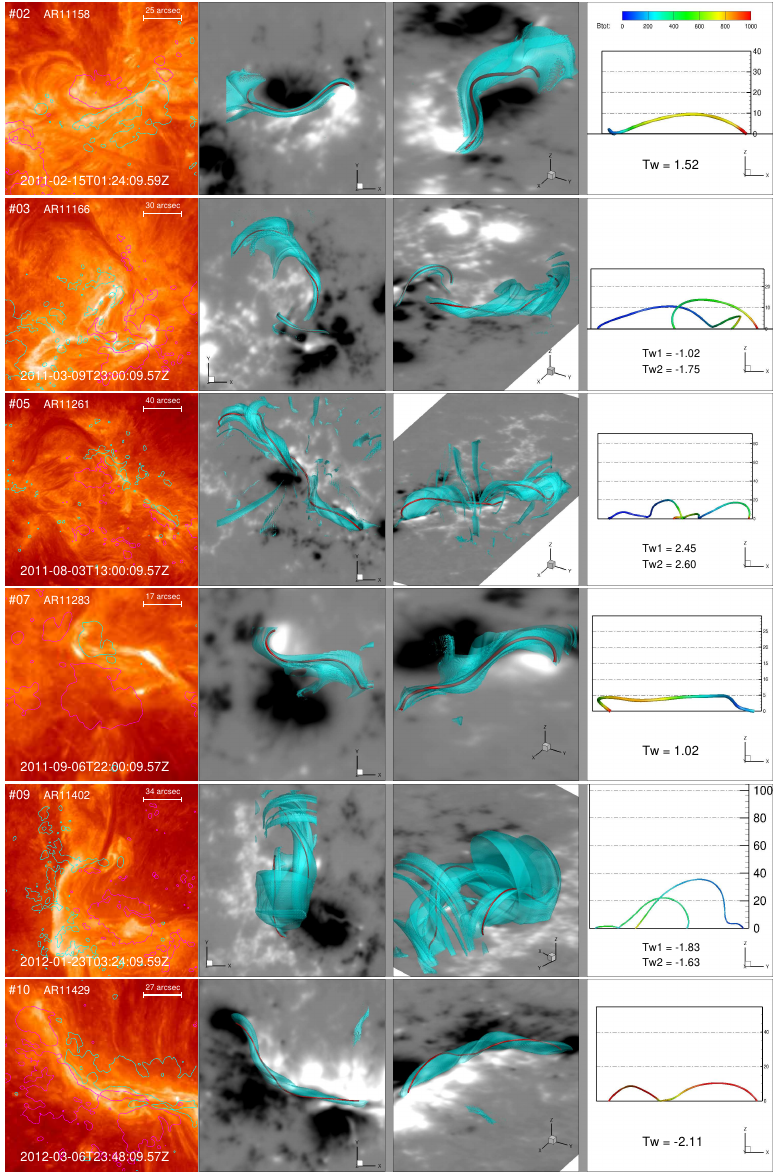}
  \caption{Configuration of the MFRs. From the left to right are
    {\SDO}/AIA~304~{\AA} image taken at the pre-flare time, the
    reconstructed MFR structure in three different views. In the AIA
    images, contour lines of photosphere $B_{z} = \pm 1000$~Gauss are
    overlaid. In the middle panels, the 3D transparent structures
    colored in cyan are iso-surface of $|T_{w}|=1$, while the thick
    red lines denote the axis of the MFRs. The background is shown
    with photosphere magnetogram (saturated at $\pm 1000$~Gauss). In
    the right panels, only the MFR's axis is shown with color denotes
    the magnetic field strength along the rope axis. For each rope
    axis, the twist number $T_{w}$ is shown. The unit of the
      $z$ axis is 1~arcsec (or 720km).}
  \label{MFR_all_1}
\end{figure*}

\begin{figure*}[htbp]
  \centering
  \includegraphics[width=0.8\textwidth]{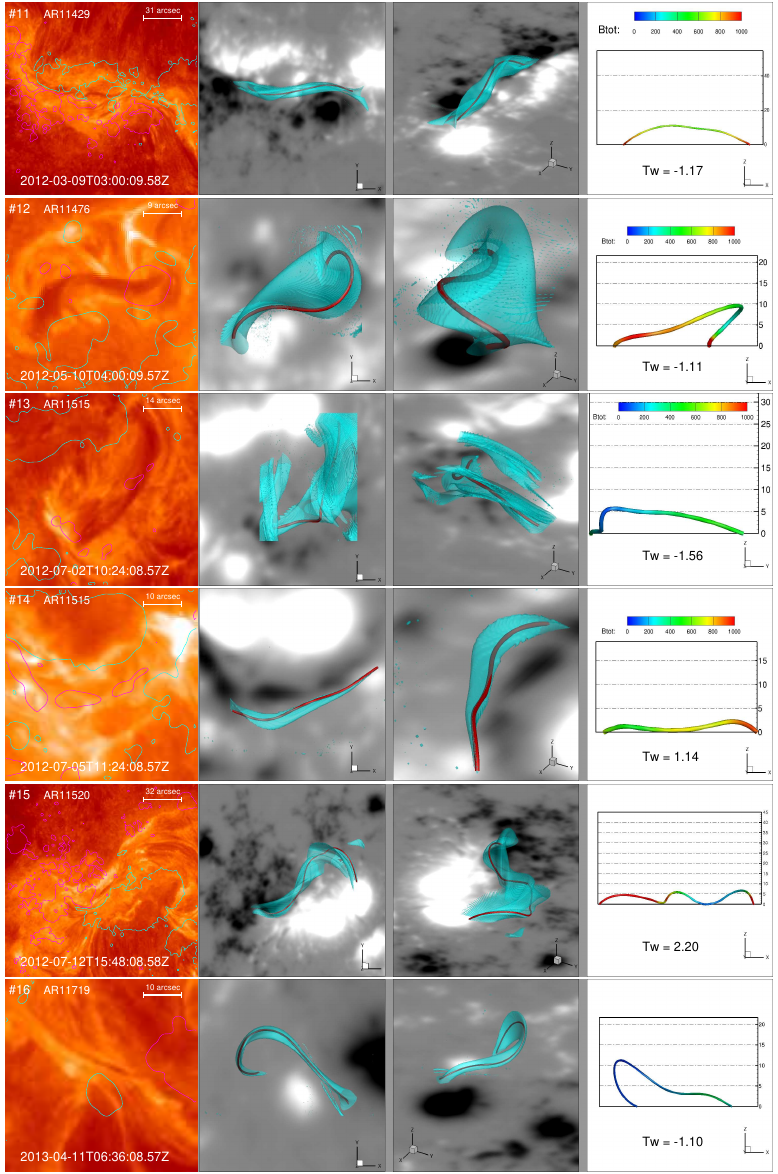}
  \caption{Same as \Fig~\ref{MFR_all_1}, but for another 6 events.}
  \label{MFR_all_2}
\end{figure*}

\begin{figure*}[htbp]
  \centering
  \includegraphics[width=0.8\textwidth]{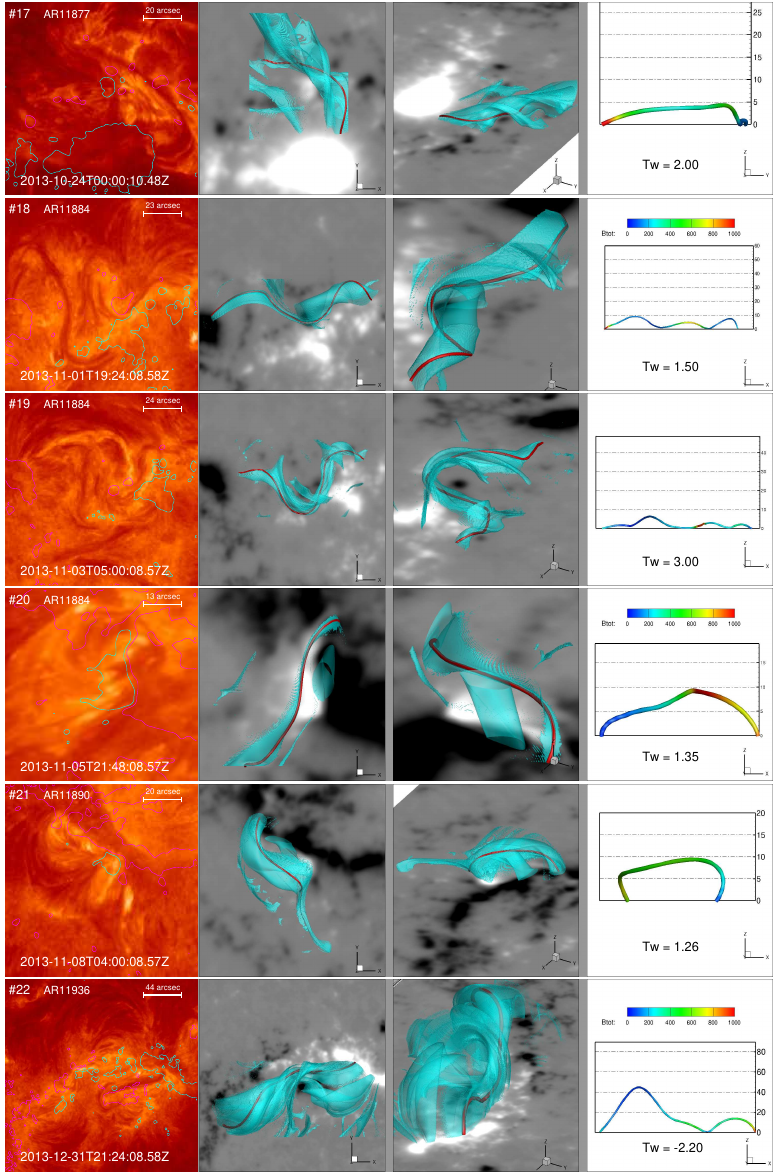}
  \caption{Same as \Fig~\ref{MFR_all_1}, but for another 6 events.}
  \label{MFR_all_3}
\end{figure*}

\begin{figure*}[htbp]
  \centering
  \includegraphics[width=0.8\textwidth]{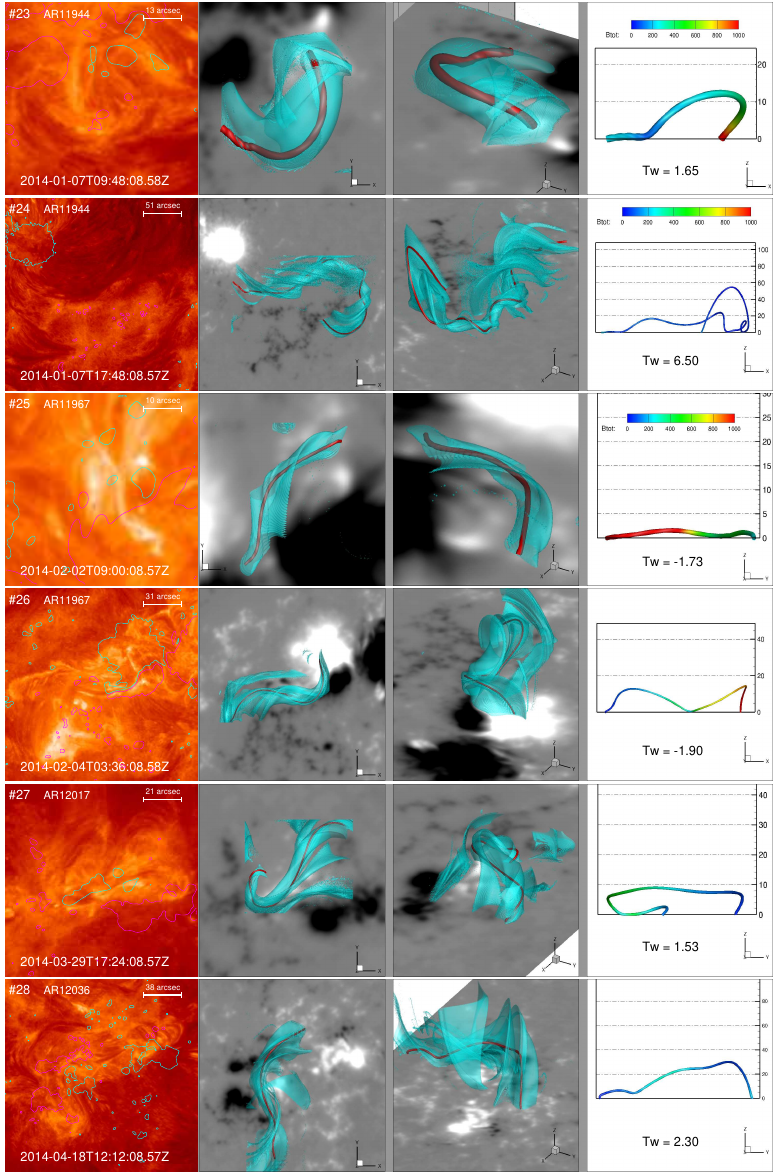}
  \caption{Same as \Fig~\ref{MFR_all_1}, but for another 6 events.}
  \label{MFR_all_4}
\end{figure*}

\begin{figure*}[htbp]
  \centering
  \includegraphics[width=0.8\textwidth]{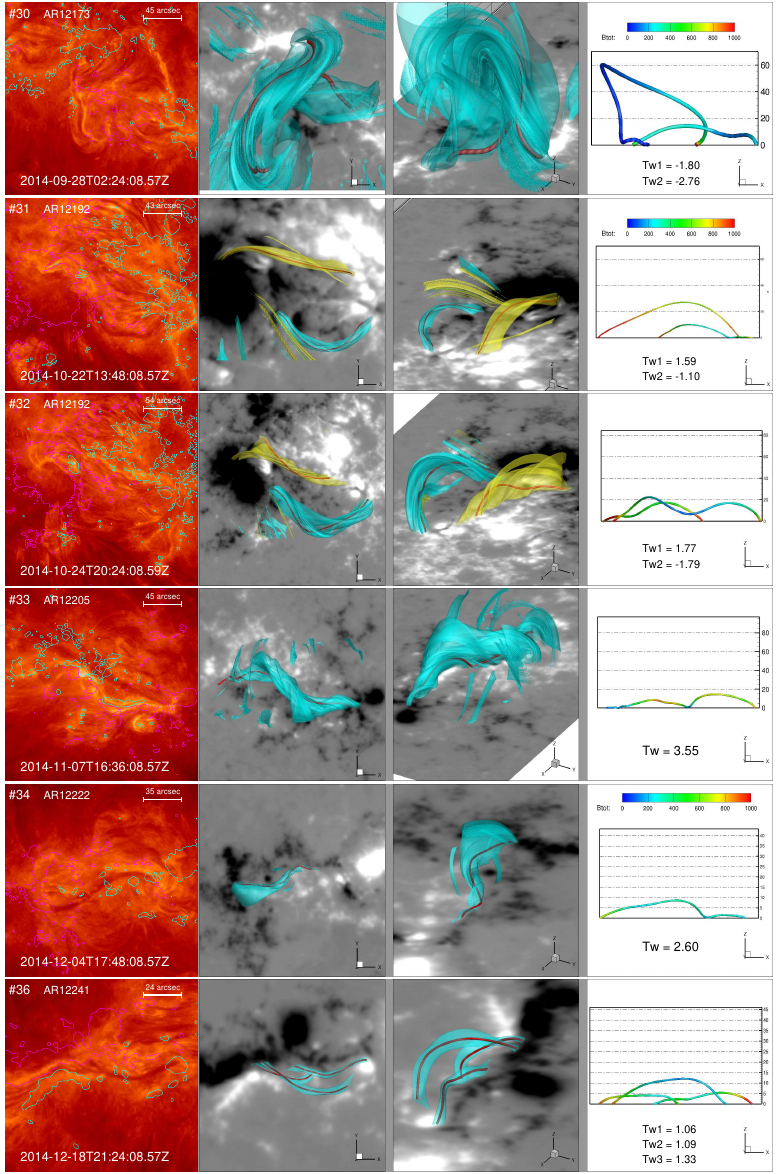}
  \caption{Same as \Fig~\ref{MFR_all_1}, but for another 6
    events. Note that in events 31 and 32, the iso-surfaces of $T_w=1$
    are colored in cyan and $T_w = -1$ colored in yellow.}
  \label{MFR_all_5}
\end{figure*}

\begin{figure*}[htbp]
  \centering
  \includegraphics[width=0.8\textwidth]{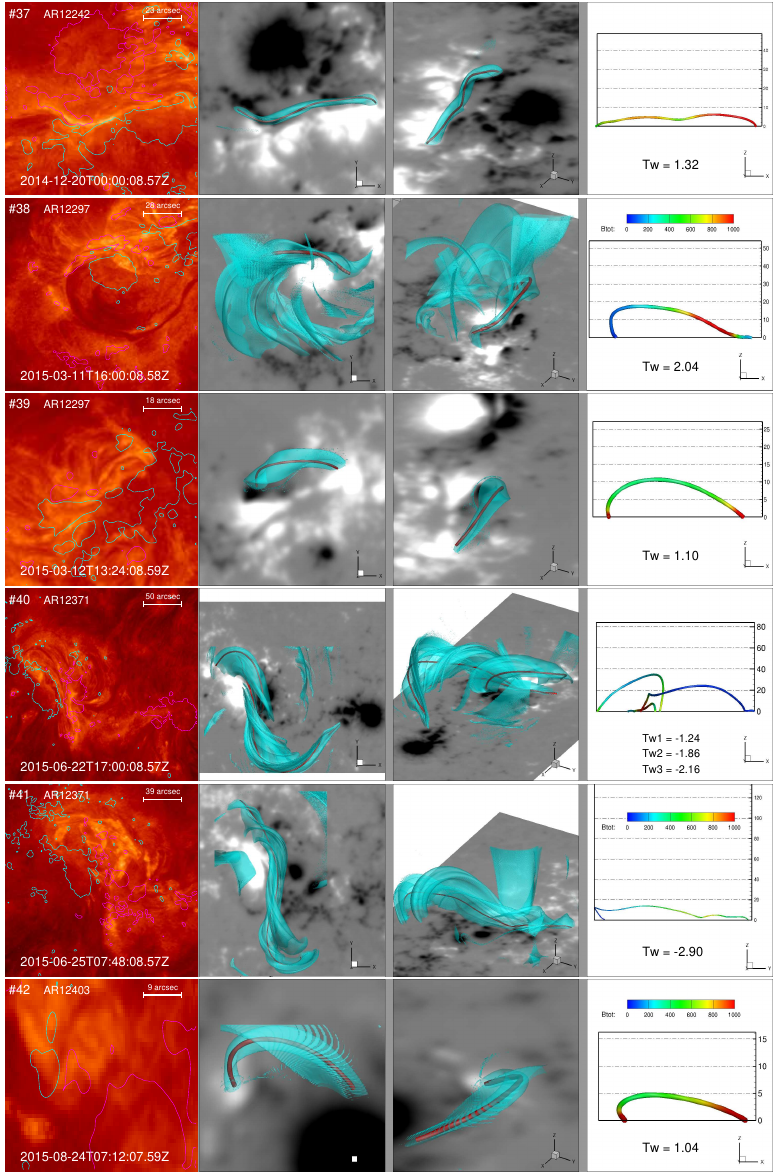}
  \caption{Same as \Fig~\ref{MFR_all_1}, but for another 6 events.}
  \label{MFR_all_6}
\end{figure*}

\begin{figure*}[htbp]
  \centering
  \includegraphics[width=0.8\textwidth]{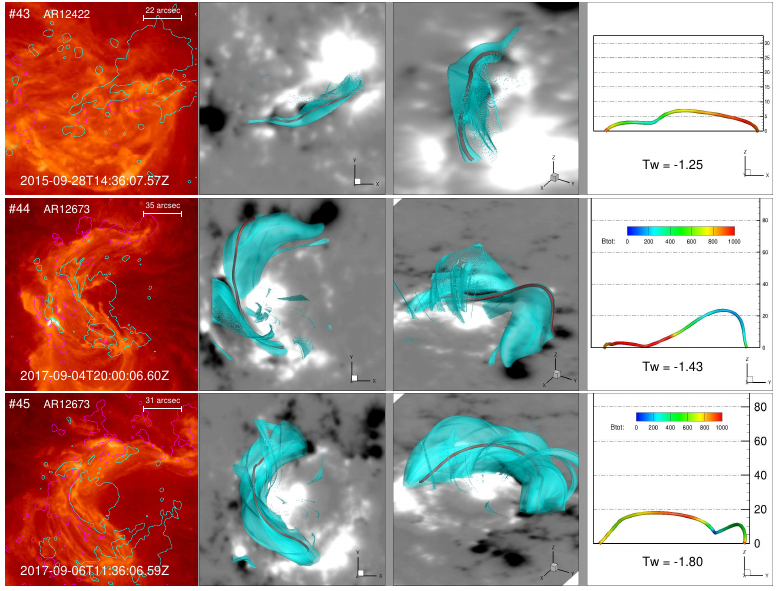}
  \caption{Same as \Fig~\ref{MFR_all_1}, but for the last 3 events.}
  \label{MFR_all_7}
\end{figure*}

\section{Results}
\label{sec:res}

\subsection{Complex Configurations}
\label{complex}
With our strict definition, only 6 of the 45 events (13\%) have no MFR
as the maximum twist number \Twm is less than 1 (although they are
very close to 1).  For the remaining 39 events, we plot the MFR
configurations in the figures from \Fig~\ref{MFR_all_1} to
\Fig~\ref{MFR_all_7}. For each event, three different angles of view
are shown and a {\SDO}/AIA-304~{\AA} image taken at the same time used
for reconstructions is present for comparison. Overall, the twist flux
forms coherently ropes and the central section of the ropes run
roughly along the main PIL of the AR, but the shapes are very
different from case to case and it is not easy to make classification
for them. Furthermore, from the morphology, very few of them show an
idealized, symmetric half-circle MFR that is often used for
theoretical study, which indicates that the complexity of coronal MFRs
is far beyond the characterization by current theoretical (or
idealized) models. The aspect ratios of the MFRs are also different,
some are rather short and thick, while some are long and thin. Many
MFRs have a serpent shape that its body touch the bottom surface one
or several times (for instance, see events 10, 15, 18, etc.),
suggesting there are bald patches along the PIL below the MFRs. Most
of the MFRs show good coherence except a few of them are disturbed by
segmented sheet-like structures of $|T_{w}|>1$ locating very close to
them. Nevertheless, the comparison of the MFR structures and the
corresponding AIA images shows good agreement between the filaments
and the MFRs. Almost for each MFR, a filament can be identified in the
AIA~304~{\AA} channel, and the spine of the filament looks co-spatial
to the axis of the MFRs.  Only in the events 16, 20, 25, and 43, there
appears to be no filament co-spatial to the MFRs. Note that in event
24, the filament is not clearly seen in AIA~304~{\AA} but a co-spatial
filament channel can be seen if check the AIA~171~{\AA} images (not
shown here).

Although the majority of the events has a single MFR, there are cases
in which we identified two MFRs in a single event (events 3, 5, 9, 30,
31, 32) or even three MFRs (events 36 and 40)\footnote{Note that for
  such multiple MFR cases, only the MFR with the largest height is
  listed in Table~\ref{tab:event_list} and used in the statistic
  analysis.}.  So in total the events with multiple-MFR configuration
accounts for 20\% of all the events (8 in 39). Actually the
multiple-MFR configuration is even more common in the ARs if
we release the restriction on only those relevant to the flare
site. For some events, for instance, number 36 and 40, there might be
no difference in magnetic topology between different flux ropes in the
multiple-MFRs configuration, and only the uneven distribution of
magnetic twist causes the complexity of the MFR system. It is worth
noting that the two MFRs in events 31 and 32, which are both confined
X-class flares from the largest region AR 12192~\citep{SunX2015,
  Jiang2016ApJ}, have inverse signs of magnetic twist. This might
provide an interesting explanation why these flares are confined, as
the inverse helicity contents of the MFRs might cancel with each other
during the flare such that there is no need for the pre-flare reserved
helicity to release out of the AR through eruption, and further
investigations on these events will be performed in future
study.

\begin{figure}[htbp]
  \centering
  \includegraphics[width=0.45\textwidth]{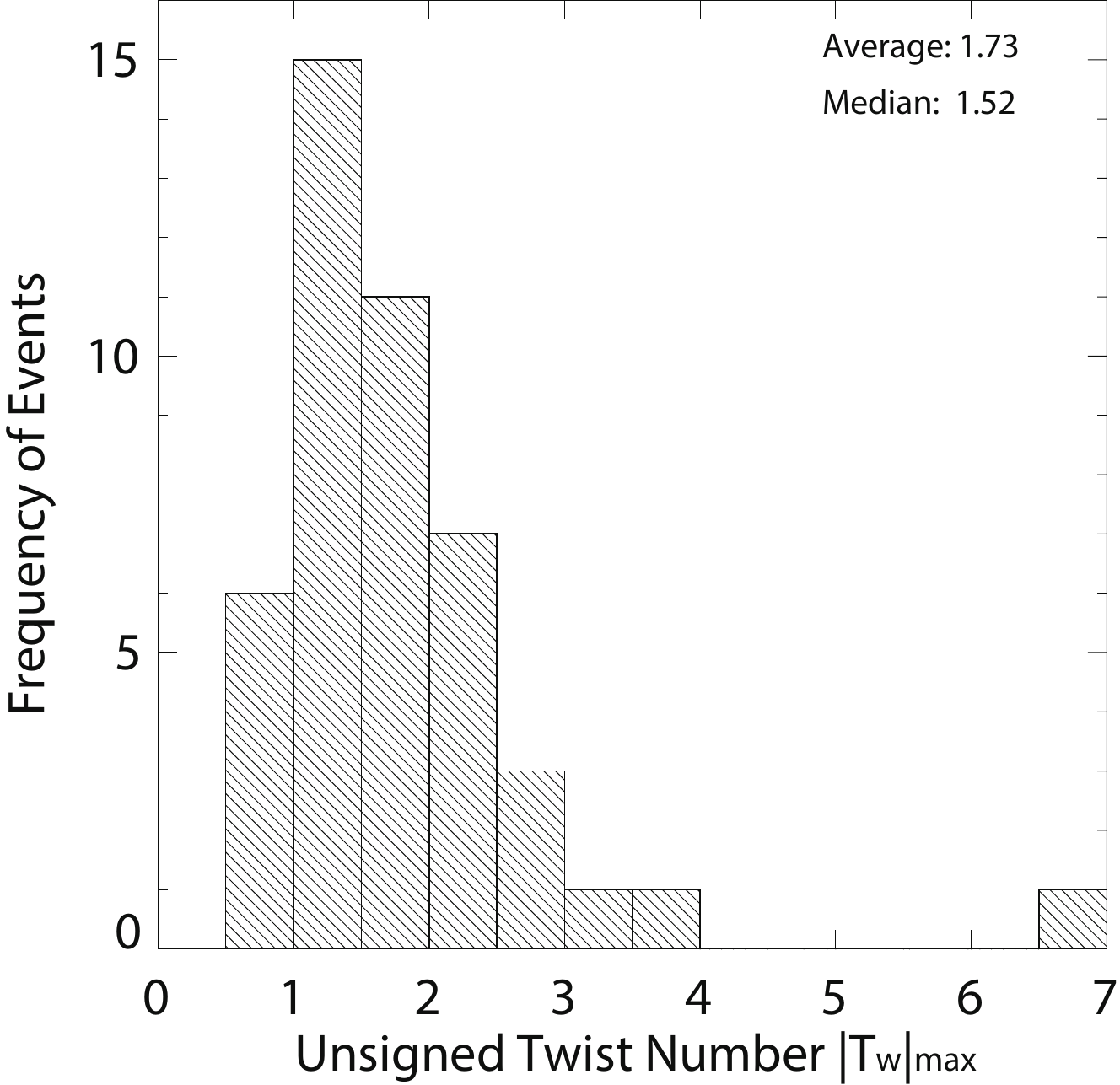}
  \caption{Histogram of the \Twm for all the MFRs. The mean and median
    values are denoted on the figure.}
  \label{tw_histogram_all}
\end{figure}

In \Fig~\ref{tw_histogram_all}, we show the distribution of the \Twm
for all the events. We find the average \Twm of 1.73 and the median
value of 1.52, and 13 events (29\%) have \Twm larger than 2. Previous
NLFFF reconstructions for pre-eruptive or quiescent filaments often
yields weakly twisted MFRs with twist number mostly below
1.5~\citep{LiuR2016, WangY2016}. Comparing with previous results, we
find that our extrapolations can reconstruct relatively high twisted
flux rope. In all cases, the largest number reaching 6.5 in the event
24, which is an X1.2 flare occurring between ARs 11944 and 11943, and
by checking the AIA images, we found that the flux rope for this event
actually corresponds to an intermediate filament channel between
the two ARs, and its flare eruption is most likely caused by KI due to the strong
magnetic twist.

\subsection{TI versus KI parameter diagram}
\begin{figure*}[htbp]
  \centering
  \includegraphics[width=0.8\textwidth]{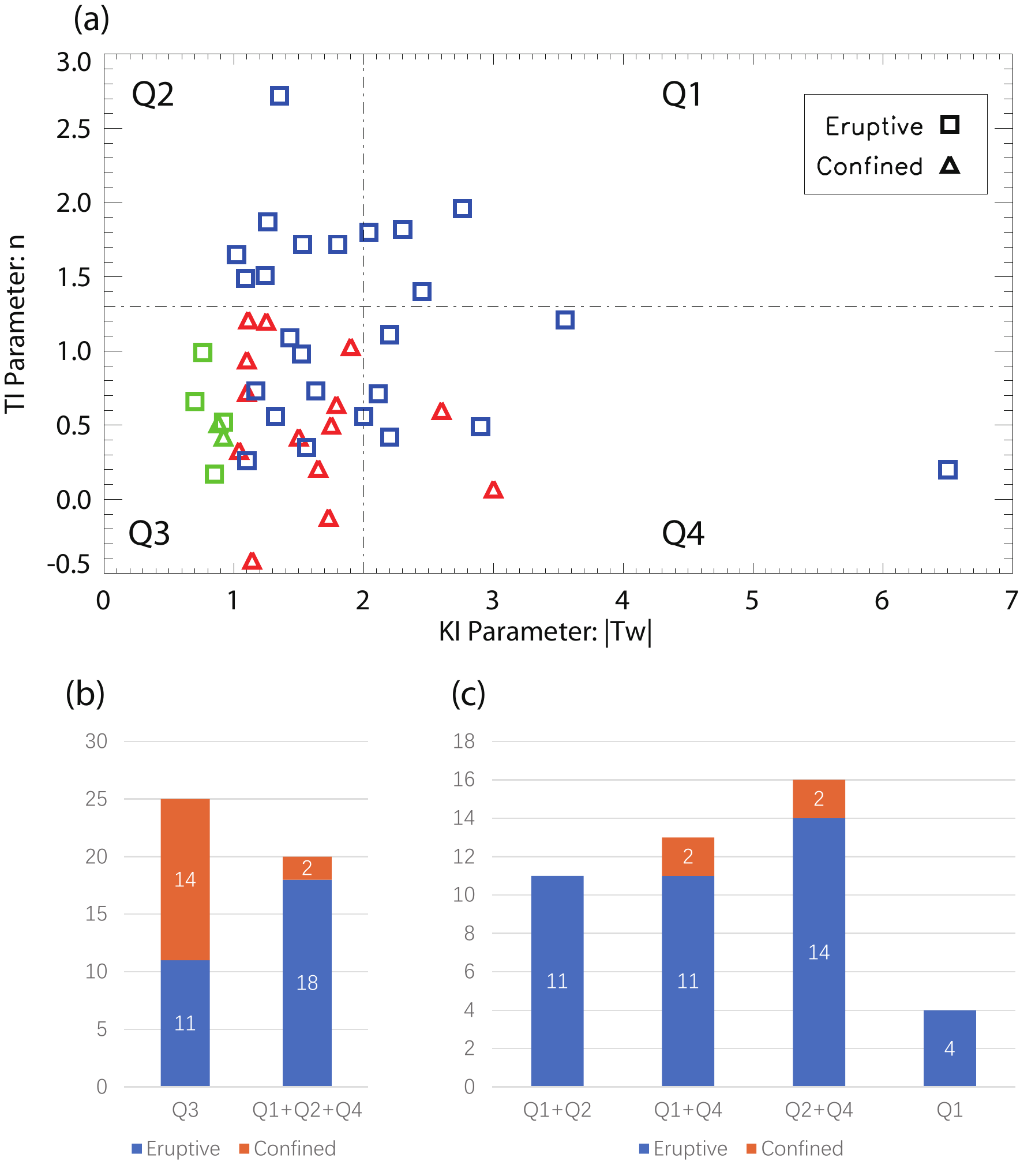}
  \caption{Scatter diagram of decay index $n$ (TI parameter) vs. \Twm
    (KI parameter) and histograms of the number of different-type
    events.  (a) The boxes denote eruptive flares and the triangles
    denote confined flares, while those in green color denotes the
    non-MFR events, i.e., \Twm is less than 1. The vertical dashed
    line marks the $|T_w|=2$, and the horizontal dashed line marks the
    $n=1.3$. Based on these two lines, the distribution of all the
    events can be divided into four quadrants, $Q1$, $Q2$, $Q3$, and
    $Q4$.  (b) Histogram for numbers of events in $Q3$ and
    $Q1+Q2+Q4$. (c) Histogram for numbers of events in $Q1+Q2$,
    $Q1+Q4$, $Q2+Q4$, and $Q1$.}
  \label{n_and_tw_histogram}
\end{figure*}

In \Fig~\ref{n_and_tw_histogram}(a), we show the scatter diagram of
decay index $n$ (TI parameter) versus the \Twm (KI parameter) for
all the 45 events. For the 6 non-MFR cases, we also calculated their
\Twm and the decay index $n$ of the field line with \Twm (shown in
green color). From the distribution of eruptive and confined flares in
the parameter space, it can be empirically identified a critical value
for $n$ and $|T_{w}|$, which are $n_{\rm crit}=1.3$ and
$|T_w|_{\rm crit}=2$, respectively, as marked on the figure. According
to these critical values, the distribution of the events falls into
four quadrants which are defined as: Q1 ($n>=1.3$ and $|T_w|>=2$), Q2
($n>=1.3$ and $|T_w|<2$), Q3 ($n<1.3$ and $|T_w|<2$), and Q4 ($n<1.3$
and $|T_w|>=2$). The histograms for events falling into different
quadrants are shown in \Fig~\ref{n_and_tw_histogram}(b) and (c). As
can be seen in the \Fig, all the events with decay index above
$n_{\rm crit}$ (i.e., in Q1+Q2) erupted. Thus $n>n_{\rm crit}$ can be
regarded as a sufficient condition for eruptive flare. For all the
events with \Twm above $|T_w|_{\rm crit}$ (i.e., in Q1+Q4), 85\%
erupted (11 in 13). This suggests that $n_{\rm crit}=1.3$ and
$|T_w|_{\rm crit}=2$ can be reliably used as the lower limits of the
threshold values for KI and TI in our statistic samples. So it is
reasonable to assume that the events fall in Q1 fulfill both TI and
KI, in Q2 fulfill only TI, in Q4 fulfill only KI, and in Q3 none of MHD instabilities is
fulfilled. Over 87\% confined events (14 in
16) reside in Q3, for which both KI and TI are not fulfilled. If
doing a prediction for the type of eruptive or confined in all the 45
events using the critical values of $n$ and $|T_{w}|$ derived from the
pre-flare field reconstructions, over 70\% are successful predicted
and the remaining 13 events include 11 eruptive ones in Q3 and 2
confined ones in Q4.

The TI threshold $n_{\rm crit}=1.3$ is close to that derived from many
theoretical models or simulations, which is $n = 1.3\sim 1.5$.
Furthermore, most of the events above $n_{\rm crit}$ have decay index
clustered within the domain of $[1.5, 1.7]$. The average value of $n$
for all the events above $n_{\rm crit}$ is $\overline{n}=1.79$ and the
standard deviation is $\sigma=0.35$. On the other hand, since the
trigger of KI depends on many parameters of MFR such as aspect ratio,
the geometry of the axis, line tying, and other details of the
configuration, there seems to be no single value or values in a narrow
region for the KI threshold. In our statistic results, the \Twm for
all eruptive events above $|T_w|_{\rm crit}$ spread out much more than
the distribution of $n$ above $n_{\rm crit}$ (the average of \Twm is
$\overline{|T_{w}|_{\rm max}} = 2.82$ and the standard deviation is
$\sigma=1.31$). This agrees with the complexity in the configurations
of the MFRs, and trigger of KI needs rather different threshold values
depending on the details of the complex configurations. With
photospheric line tying included, the minimum threshold from theory is
a winding of the field lines about the rope axis by $1.25$
turns \citep{Hood1981}. Our statistic results suggests
that the KI threshold is much larger than this minimum
value, but the lower limit $|T_w|_{\rm crit}=2$ is comparable with estimations of
 winding number from numerical models of solar flux ropes~\citep[e.g., 1.75 in ][]{Torok2004} and \citep[1.875 in][]{Fan2003}.

If we assume that all the eruptive events with $n \ge n_{\rm crit}$
are triggered by TI (i.e., events in Q1+Q2), and meanwhile all the
eruptive events with $|T_{w}| \ge |T_w|_{\rm crit}$ are caused by KI (i.e., events
in Q1+Q4 except the two confined ones), their numbers are equal (see
\Fig~\ref{n_and_tw_histogram}c). This indicates that KI plays an
equally important role as TI in triggering eruptions. On the other hand,
the total number of the eruptive events in Q2+Q4, that is triggered by either TI or KI,
is 14. This is much larger than that in Q1 (events
fulfill both TI and KI) which have only 4 events, suggesting that TI
and KI do not necessarily to be fulfilled simultaneously to trigger
an eruption.

In summary, our analysis suggests that either of the two parameters,
$n$ and $|T_{w}|_{\rm max}$, provides a strong constraint on the
eruptiveness of major flares, i.e., if $n \ge n_{\rm crit}$ or
$|T_{w}|_{\rm max} \ge |T_w|_{\rm crit}$, the event has a very high
possibility (90\% in our studied samples) of successful eruption. For
the remaining events with both $n<n_{\rm crit}$ and
$|T_{w}|_{\rm max}<|T_w|_{\rm crit}$, they can either be eruptive or
confined. Furthermore, their distribution in this domain (Q3) appears
rather random between the eruptive and confined ones, suggesting the
$T_w$ and $n$, or, KI and TI, cannot differentiate the types, which
hints that magnetic reconnection rather than ideal MHD instability of
MFR is the flare trigger. Their total number is 25, i.e., 56\% of the
all events. Interestingly, there are two cases in which $n$ is
negative, because the MFR is too low that the overlying flux initially
increases rather than decreases. Finally, as noted in
Section~\ref{complex} there are 4 events having no filament co-spatial
with the reconstructed MFRs, which are event 16 (eruptive in Q3), 20
(eruptive in Q2 with the largest $n$), 25 (confined in Q3), and 43
(confined in Q3). If excluding those events, our findings remain valid
with only some of the percentages changing slightly.

\section{Discussions and Conclusions}
\label{sec:concl}
In this paper, we carried out a survey of the MFRs existing
immediately before major solar flares (generally above GOES M5 class)
using a coronal magnetic field reconstruction method with {\SDO}/HMI
vector magnetograms. By analyzing the configurations and two key
parameters, which are decay index and the maximum twist number in the
MFR, for ideal MHD instabilities of MFR, we have the following key
findings.

1. In consistence with many previous case
studies~\citep[e.g.,][]{Rust2003, Gibson2006, Canou2009, Green2009,
  Yeates2009, Canou2010, Savcheva2015, Su2015, Yurchyshyn2015,
  ZouP2019}, MFRs generally exists prior to major solar flares. With a
rigorous definition, there are over 90\% of the studied events have
well-defined MFRs in the flare site, i.e., a coherent group of
magnetic field lines with twist above one turn and the field line
possessing the peak value of twist as being the rope axis.  The rest
10\% events also have MFR-like structures as their magnetic twist
numbers are very close to one. The maximum twist numbers in the MFRs
have an average value of 1.73 for all the events, which is
systematically higher than that from other reconstruction
methods~\citep[e.g.,][]{WangY2016, LiuR2016, JingJ2018}.  Most of the
MFRs have corresponding filaments or filament channels as seen in
{\SDO}/AIA 304 {\AA} observations.

2. It is the first time that all the pre-flare MFRs of such large
sample are presented with 3D configuration.  The reconstructed MFRs
demonstrated much more complex configurations in details than
idealized models of MFR that are often used in theoretical
investigations or numerical simulations~\citep[e.g.,][]{Titov1999,
  Torok2005, Aulanier2010, MeiZ2018}. Furthermore, multiple MFRs are
found in 20\% of the events, and in a few cases (in the AR~12192), the
MFRs can even have inverse signs of magnetic twist. This might provide
a new way of explaining the confinement of the flares from the point
of view of magnetic helicity and further investigation is required.

3. The parameter diagram formed by the twist number of the MFR axis
and its decay index suggests a lower limit for TI and KI thresholds,
which are $n_{\rm crit}=1.3$ and $|T_{w}|_{\rm crit}=2$,
respectively. All the events above the $n_{\rm crit}$, and nearly 90\%
of the events above the $|T_{w}|_{\rm crit}$ erupted. The eruptive
events above the TI threshold have an average decay index with a small
deviation of $1.79\pm 0.35$, which is close to many theoretical
derived TI thresholds, although the reconstructed MFRs are much more
complex than the theoretical ones. On the other hand, the values for
KI threshold spread out in a wider domain of $2.83\pm 1.31$. From this
result, an important argument can be made: the KI are more sensitive
to the details of the MFR itself while the TI depends mainly on
decaying speed of the strapping field overlying the MFR~\footnote{Of
  course, regarding the TI, there must be a MFR before the flare,
  otherwise the TI is meaningless. However, this requirement seems to
  be often ignored in application of the TI theory to
  observations. This also might explains why there is a so-called
  `failed torus' regime as reported in \citet{Myers2015} because in
  that regime the magnetic twist numbers are mostly less than one, and
  thus by a strict definition, the MFR barely exists.}.

4. Our results show significant difference from a previous similar
study by~\citet{JingJ2018}. In their findings, the lower limit of TI
threshold is $n_{\rm crit}=\sim 0.75$, and the KI seems to play no
role in differentiating the eruptive and confined events. On the
contrary, our results show that the \textbf{TI} threshold
($n_{\rm crit}$) is much higher, and furthermore, KI is equally
important as TI in producing eruption. Such difference of our results
from \citet{JingJ2018}'s can be attributed to many factors, and we
suspect that the leading one is that different coronal field
reconstruction methods were used (the Wiegelmann's code was used in
\citet{JingJ2018}). The inconsistence between different reconstruction
codes applied to the real data has been extensively
reported~\citep[e.g.,][]{DeRosa2009, Regnier2013, Aschwanden2014,
  DuanA2017, Wiegelmann2017}, even though they can provide rather
consistent results in some benchmark tests using idealized or
artificial magnetograms. Furthermore, the computational method of
decay index at the MFR's axis and the using of the maximum $|T_{w}|$
in the MFR as the KI parameter are also different from the analysis
method in
\citet{JingJ2018}. 
Since the results in this paper is more consistent with theoretical
studies, we thus suggest that it might be more relevant to follow the
approach taken here in future application of TI and KI to
reconstructed coronal field. On the one hand, our definition of MFR is
much stricter and the maximum of twist number of MFR is more sensitive
than the average twist that includes also the contribution of the flux
with $|T_{w}|<1$, as such weakly twisted flux is not relevant to the
MFRs. On the other hand, computing the oblique decay index is more
relevant to the eruption at the early phase, i.e., its initiation due
to the complex, non-symmetric magnetic environments.

5. Comparing to the eruptive events fulfilled either TI or KI, the
events fulfilled both are minor. This suggests that TI and KI do not
necessarily to be fulfilled at the same time to trigger an eruption.
This is reasonable as on the one hand, the presence of a well-defined
MFR (i.e., with magnetic twist number above 1) will erupt once it runs
into the TI regime, and on the other hand a initially torus-stable MFR
can be lifted to TI threshold through the kinking deformation of the
MFR.

6. The events with both decay index and twist degree below our derived
lower limits of TI and KI accounts for 56\% (and among them 44\% are
eruptive). For these events, we conjecture that the ideal MHD
instabilities might not be able to trigger the flare, and the
non-ideal MHD mechanism, i.e., magnetic reconnection, should play an
more important role, in particular, in producing the successful
eruption. Recently, Zou et al. (2019, in press) have carried out a
statistic survey for all the AR filament eruptions from 2011 to 2017
which produced fast CMEs (above 800~km~s$^{-1}$) and found that over
60\% of AR filaments are more likely triggered by reconnection rather
than the ideal MHD instability. Our percentage of 56\% is close to
their result, which further indicates that magnetic reconnection
(e.g., tether cutting and breakout) plays an important role (at least,
equal to the ideal MHD instabilities) in the triggering of major
flares and eruptions. Thus for understanding the mechanism of these
flares, a detailed analysis of the magnetic topology that is favorable
for reconnection is required. In particular, more attention should be
paid on the eruptive events in the regime to see how the reconnection
can break through the strong confinement of the overlying field, which
will be left in future studies.

\acknowledgments

This work is jointly supported by National Natural Science Foundation
of China (41604140) and the startup funding (74110-18841214) from Sun Yat-sen University.  C.J. acknowledges support by National Natural Science
Foundation of China (41822404, 41731067, 41574170, 41531073).
Data from observations are courtesy of NASA {SDO}/AIA and
the HMI science teams. Special thanks to our anonymous reviewer for valuable
suggestions that helped improve the paper.


\end{CJK*}
\end{document}